\documentclass[amsmath,amssymb,superscriptaddress]{revtex4}

\usepackage{graphicx}
\usepackage{dcolumn}

\def\be{\begin{equation}}
\def\ee{\end{equation}}
\def\beq{\begin{eqnarray}}
\def\eeq{\end{eqnarray}}

\usepackage{bm}
\usepackage{graphicx}
\usepackage{dcolumn}
\usepackage{amsmath}
\usepackage[latin1]{inputenc}
\usepackage{graphicx, psfrag}
\usepackage{amssymb}
\usepackage[colorlinks=true, citecolor=blue, urlcolor = blue, linkcolor= red, bookmarks=true]{hyperref}
\usepackage{float}
\usepackage{amsmath}
\usepackage{amsfonts}
\usepackage{dcolumn}
\usepackage{hyperref}
\usepackage{subfigure}
\usepackage{pgfplots}
\usepackage{epstopdf}
\usepackage{booktabs}

\begin{document}


\title{Anisotropic Stars in $4D$ Einstein-Gauss-Bonnet Gravity}



\author{Takol Tangphati} \email[]{takoltang@gmail.com}
\affiliation{Department of Physics, Faculty of Science, Chulalongkorn University, \\Bangkok 10330, Thailand,}

\author{Anirudh Pradhan} \email[]{pradhan.anirudh@gmail.com}
\affiliation{Department of Mathematics, Institute of Applied Sciences and Humanities, GLA University, Mathura-281 406, Uttar Pradesh, India}

\author{Ayan Banerjee} \email[]{ayanbanerjeemath@gmail.com}
\affiliation{Astrophysics and Cosmology Research Unit, School of Mathematics, Statistics and Computer Science, University of KwaZulu--Natal, Private Bag X54001, Durban 4000, South Africa}

\author{Grigoris Panotopoulos} \email[]{grigorios.panotopoulos@tecnico.ulisboa.pt}
\affiliation{Centro de Astrof{\'i}sica e Gravita{\c c}{\~a}o-CENTRA, Instituto Superior T{\'e}cnico-IST, Universidade de Lisboa-UL, Av. Rovisco Pais, 1049-001 Lisboa, Portugal}
\affiliation{Departamento de Ciencias F{\'i}sicas, Universidad de la Frontera, Avenida Francisco Salazar, 01145 Temuco, Chile}


\date{\today}

\begin{abstract}
The current trend concerning dense matter physics at sufficiently high densities and low temperatures is expected to behave as a degenerate Fermi gas of quarks forming Cooper pairs, namely a color superconductor, in the core of compact objects. In this context, we study the anisotropy of quark stars (QSs) assuming the internal composition to be comprised of homogeneous, charge neutral 3-flavor interacting quark matter with $\mathcal{O}(m_s^4)$ corrections. Using the equation of state (EoS) with the Tolmann-Oppenheimer-Volkoff (TOV) structure equations, we perform numerical calculation for quark stars and determine the maximum mass-radius relation in the context of $4D$ Einstein-Gauss-Bonnet (EGB) gravity. In particular, we consider the effects of Gauss-Bonnet (GB) coupling
constant on the diagrams related to mass-radius $(M-R)$ relation and the mass-central mass density $(M-\rho_c)$ relation of QSs. We pay particular attention to the influence of the anisotropy in the equilibrium and stability of strange stars. We also study the other properties of QSs related to compactness and binding energy. Interestingly,  our result provides circumstantial evidence in favor of super-massive pulsars in $4D$ EGB gravity.

\end{abstract}


\maketitle


\section{Introduction}\label{intro:Sec}


In recent years a number of modified gravity theories have been proposed to address several shortcomings in order to achieve a self-consistent picture of the observed Universe.  In particular, higher curvature gravity theories appear in an effective level. Among the higher curvature gravities, so-called Einstein-Gauss-Bonnet (EGB) gravity that emerges as a low energy effective  action of  heterotic string theory \citep{Zwiebach:1985uq}. The EGB theories have some special characteristics that contain quadratic powers of the curvature and are free of ghost. In addition, the Gauss-Bonnet (GB) term yields non-trivial dynamics in $D > 4$. Inclusion of this term in the action leads to a variety of new solutions, see \cite{Charmousis:2008kc}. For example, black hole solution was first obtained by Boulware and Deser \cite{Boulware:1985wk}  in $D\geq 5$-dimensional EGB gravitational theory. However several other types of solutions exist in higher curvature gravity (see e.g.
\cite{Ghosh:2016ddh,Rubiera-Garcia:2015yga,Giacomini:2015dwa,Aranguiz:2015voa,Xu:2015eia}). Likewise, the spherical gravitational collapse of inhomogeneous dust has been explored in 
\cite{Jhingan:2010zz,Maeda:2006pm,Zhou:2011vy,Abbas:2015sja}.  In \cite{Tangphati:2021tcy}, authors have studied anisotropic quark stars in the same context. 

\smallskip

Nevertheless, in $4D$ the GB term is a topological invariant, hence it does not contribute to the gravitational dynamics. However, a novel 4$D$ EGB gravity theory was recently proposed by Glavan and Lin \cite{Glavan:2019inb}, who showed that there are nontrivial effects from the GB term even in four dimensions. Their basic idea was to rescale the GB coupling constant by $\alpha \to \alpha/(D -4)$ 
in D-dimensional space-time, and then taking the limit $D \to 4$. As a consequence of this limit, one can bypass the conclusions of Lovelock's theorem and avoid the Ostrogradsky instability \cite{Woodard:2015zca}.
The resultant theory is now dubbed as the novel 4$D$ EGB theory, and it has stimulated a lot of studies as well as doubts. Since then, a number of interesting solutions have been found in this theory, such as black hole solutions including their physical properties 
\cite{Ghosh:2020syx,Konoplya:2020juj,Singh:2020xju,HosseiniMansoori:2020yfj,Singh:2020nwo,Wei:2020poh,Yang:2020jno}, charged black hole \cite{Fernandes:2020rpa,Zhang:2020sjh}, black holes coupled to magnetic charge and  nonlinear electrodynamics \cite{Jusufi:2020qyw,Abdujabbarov:2020jla,Jafarzade:2020ilt}. Beside that, deflection of light by black holes 
\cite{Islam:2020xmy,Jin:2020emq,Kumar:2020sag},  quasi-normal modes \cite{Churilova:2020aca,Mishra:2020gce,Aragon:2020qdc}
and shadow cast by black holes \cite{Konoplya:2020bxa,Guo:2020zmf,Zeng:2020dco} have also been investigated. Moreover, Morris-Thorne-like wormholes and thin-shell wormholes have been investigated in  \cite{Jusufi:2020yus,Liu:2020yhu}.  
In Ref. \cite{Cao:2021nng}, it was shown that in order to have
a well defined linearized theory the metric should be (locally) conformally flat. Keeping in view the importance of this theory several works have done to constraint the new re-scaled Gauss-Bonnet parameter from observational data, see Refs. \cite{Wang:2021kuw,Feng:2020duo,Clifton:2020xhc} and so on. 

\smallskip

On the other hand, neutron stars (NSs) are dense, compact astrophysical objects which  are strictly correlated with the equation of state (EoS), i.e. the relation between density and pressure in its interior. Unfortunately, the composition and inner structure of a NS is not known in detail. Due to this lack of knowledge, physicists have to speculate analysing different models of 
NSs based on various EoS, each one corresponding to a different mass-radius ($M - R$) profile. Consequently, the knowledge of those parameters provides us with significant information about the evolutionary history of NSs. Thus, several models for EoS are predicted by different research groups, but only few of 
them are viable and realistic, capable of describing matter inside compact objects. 
\smallskip

The most reliable observational constraints on the properties of the maximum-mass NSs  come from the discoveries of radio pulsars and some other accretion-powered X-ray sources. Observed NS masses close to 2 $M_{\odot}$ \cite{Demorest:2010bx,Antoniadis:2013pzd} with their radii are
about $R \lesssim (11 \sim 14)$ Km \cite{Steiner:2010fz}  have provided a strong constraint on the EoS of matter. But the dynamical evolution of such violent events and the behavior of matter at high density (above the nuclear density:
$\rho_{\text{nuc}} = 2.8 \times 10^{14}$ $g/cm^3$) and high temperature (T $\sim$ 10 MeV) are still unknown to us. As a consequence, the concept of quark has been proposed by Gell-Mann and Zweig  \cite{GellMann:1964nj,Zweig:1964jf}, and quark matter is assumed to be absolutely stable and the true ground state of hadronic matter \cite{Witten:1984rs,Bodmer:1971we}. Thus, compact stars 
partially or totally made of  \textit{quark matter} hypothesis are called quark stars (QSs), and they were postulated to exist in the 70's. On the other hand, the study of dense quark matter draws much attention due to the recent development in understanding of color superconductivity (see Refs. \cite{Rajagopal:2000wf,Alford:2001dt,Hess:2011qw} for reviews). Thus, the existence of the CFL phase can enhance the possibility of the existence of a pure stable quark star have been obtained in \cite{Singh:2020bdv,Flores:2017hpb,Roupas:2020nua,Bogadi:2020sjy}.

\smallskip

In a series of recent papers, a new interest has grown on the problem of finding compact objects made of anisotropic matter. Anisotropic matter - in a spherically symmetric context means that the pressure along the radial direction, $P_r$, is different than the transverse pressure, $P_T$. Many efforts have been done to address challenges concerning anisotropic matter as interior models for spherical symmetry fluid sphere, such as boson stars \cite{Schunck}, 
gravastars \cite{Cattoen:2005he}, neutron stars \cite{Heintzmann} and so on. The main causes could be the realization of physics under extreme conditions at the  core of NSs. In Ref. \cite{Lema}
Lema$\hat{\text{i}}$te first pointed out the conception of a relativistic anisotropic sphere. After that Bowers and Liang \cite{bowers} boosted the whole idea considering the pressure anisotropy on NSs. They showed that for \textit{arbitrary large anisotropy} there is no limiting mass for NS. However, Ruderman \cite{Ruderman} first observed that nuclear matter tends to become anisotropic at very high densities of order $10^{15}$  g/cm$^3$. Herrera and Santos \cite{Herrera:1997plx} conjectured that anisotropic stars could exist in a strong gravity region.

\smallskip

The study of anisotropic stars within General Relativity (GR) is important in current research. The effects of anisotropy on slowly rotating stars have been studied in \cite{Silva:2014fca}. In \cite{Harko:2011nu}, authors have studied the two-fluid model that can be described as an effective single anisotropic fluid. Moreover, the pressure anisotropic has an impact on NS properties, such as the mass-radius relation and stability \cite{Horvat:2010xf}. 
In \cite{Herrera:2007kz} the authors studied all static spherically symmetric anisotropic solutions. Consequently, higher dimensional anisotropic compact star \cite{Bhar:2014tqa}, strange stars in 
Krori-Barua space-time \cite{Rahaman:2011cw,Kalam:2013dfa} and different kinds of exact solution of Einstein's field equations can also be found in \cite{Mak:2001eb,Maurya:2017uyk,Pretel:2020xuo}. Not surprisingly,
anisotropic solution has also been considered in modified gravity theories, see Refs. \cite{Maurya:2019msr,Maurya:2019sfm,Rahaman:2020dgv,Deb:2017rhc} for more.  

\smallskip

Above all, connect the evidence to the argument that anisotropic matter might be a reasonable candidate under extreme physical conditions. Thus, we propose to perform a general  analysis of physical viability and stability of anisotropic solutions for QSs within the framework of 4$D$ EGB gravity. This paper is organised as follows: In Sec. \ref{sec2}, we derive the field equations for $4D$ EGB gravity. We describe the problem and derive the TOV equations for static, spherically symmetric stellar metric. In Section \ref{sec3} we present an overview of 
a QCD motivated EoS. In Section \ref{sec4}, we present the numerical analysis for the QSs.  In Section \ref{sec5}, we demonstrate the effects of modified gravity and fluid anisotropy on the $M-R$ relation, and on the internal structure of the QSs. Here, we make a comparison between GR and the $4D$ EGB theory, and discuss the effect of Gauss-Bonnet coupling constant on $M-R$ relation. We present the dynamical stability as well as other properties of QSs, such as compactness and binding energy in Section \ref{sec6}. Finally, we summarize and conclude our work in section \ref{sec7}.

 \section{Stellar modelling in 4$D$ EGB gravity} \label{sec2}
 In this section, we give a brief review of the $4D$ EGB gravity, and derive the field equations as well as the structure equations for stellar interior solutions. However, a number of recent studies have shown that the limiting procedure cannot properly describe topologically nontrivial solutions \cite{Kobayashi:2020wqy,Bonifacio:2020vbk}. Various works have been initiated and shown inconsistencies in the proposed approach. In \cite{Gurses:2020ofy}, the authors pointed out that the resulting equations of motion can be decomposed into two separate parts, one being proportional to the $(D-4)$ factor, and the other being proportional to the Weyl tensor, does not have that factor, see also \cite{Ai:2020peo,Mahapatra:2020rds} for more details. This indicates that the second term does not have a smooth limit for $D \to 4$, making the equation of motion a divergent one \cite{Hohmann:2020cor}. Besides, it has been shown that $4D$ EGB gravity is not equivalent to the special Horndeski gravity for the general case \cite{Tian:2020nzb}.
 
 As a result, there are several attempts in the literature to resolve those problems, and to obtain a consistent $4D$ theory preserving certain features of the original idea. One possible approach is the compactification of $D$-dimensional EGB gravity on a maximally symmetric space of $(D-4)$ dimensions, namely, the Kaluza-Klein reduction ansatz \cite{Kobayashi:2020wqy,Ma:2020ufk,Lu:2020iav} and the resulting theories belong to the family of Horndeski gravity \cite{Horndeski}. In Refs. \cite{Fernandes:2020nbq,Hennigar:2020lsl}, a well defined $D \to 4$ limit of Gauss-Bonnet Gravity is obtained generalizing a method employed by Mann and Ross \cite{Mann:1992ar} for Einstein gravity in $D \to 2$ dimensions. In fact, one may find another consistent approach in $4D$ by breaking the invariance of the theory under diffeomorphisms \cite{Aoki:2020lig} (see also  \cite{Aoki:2020iwm} in the cosmological context).
 
 Nevertheless, it was pointed out in Ref. \cite{Fernandes:2020nbq} that the on-shell action of the regularized $4D$ EGB theory has the same form as the action of the original $4D$ EGB theory. In fact, black hole solutions in the novel  $4D$ EGB gravity \cite{Glavan:2019inb} still remain valid also in those regularized theories,  see \cite{Hennigar:2020lsl,Casalino:2020kbt}. Moreover, it has been pointed out that static, spherically symmetric $4D$ solutions still remain valid in those regularized theories \cite{Banerjee:2020yhu}, see also \cite{Lin:2020kqe} for cylindrically symmetric spacetime. As a result, the spherically symmetric solution for compact stars itself is meaningful and worthy of study.
 
In the present work, we start by deriving the equations of motion based on the novel $4D$ EGB gravity, and  the action in $D$-dimensions can be written down as
\begin{eqnarray}\label{action}
 S = \frac{c^4}{16\pi G_D} \int_{\mathcal{M}}d^D x \sqrt{-g} \left(R + \frac{\alpha}{D-4} \mathcal{L}_{\text{GB}} \right) + \mathcal{S}_{\text{matter}},
\end{eqnarray}
where  $R$ is the Ricci scalar and  $g$ is the determinant of the metric tensor $g_{\mu\nu}$. The GB coupling constant $\alpha$ has dimension of $[length]^2$,  and $\mathcal{L}_{\text{GB}}$ is the GB term, defined by 
\begin{equation}
\mathcal{L}_{\text{GB}} \equiv R^{\mu\nu\rho\sigma} R_{\mu\nu\rho\sigma}- 4 R^{\mu\nu}R_{\mu\nu}+ R^2\label{GB}.
\end{equation}

In Eq. (\ref{action}), the $\mathcal{S}_m$ is the anisotropic distribution of matter Lagrangian that depends only on the metric tensor components $g_{\mu\nu}$, and not on its derivatives. The variation of (\ref{action}) with respect to the metric tensor, $g_{\mu \nu}$, leads to the following field equations
\begin{equation}\label{GBeq}
G_{\mu\nu}+\frac{\alpha}{D-4} H_{\mu\nu}= 8 \pi T_{\mu\nu},
\end{equation}
where the Einstein tensor $G_{\mu\nu}$ and Lanczos tensor
$H_{\mu\nu}$ are defined by
\begin{eqnarray}
&& G_{\mu\nu} \equiv R_{\mu\nu}-\frac{1}{2}R~ g_{\mu\nu},\nonumber\\
&& H_{\mu\nu} \equiv 2\Bigr( R R_{\mu\nu}-2R_{\mu\sigma} {R}{^\sigma}_{\nu} -2 R_{\mu\sigma\nu\rho}{R}^{\sigma\rho} - R_{\mu\sigma\rho\delta}{R}^{\sigma\rho\delta}{_\nu}\Bigl)- \frac{1}{2}~g_{\mu\nu}~\mathcal{L}_{\text{GB}},\label{FieldEq}\\
\end{eqnarray}
while the energy-momentum tensor is given by
\begin{equation}
T_{\mu\nu}= -\frac{2}{\sqrt{-g}}\frac{\delta\left(\sqrt{-g}\mathcal{L}_m\right)}{\delta g^{\mu\nu}},
\end{equation}
where $T_{\mu\nu}$ is the energy-momentum tensor of matter field, with $R$ being the Ricci scalar, $R_{\mu\nu}$ the Ricci tensor and $R_{\mu\sigma\nu\rho}$ being the Riemann tensor, respectively.
Note that in action (\ref{action}) we have rescaled the coupling constant $\alpha \to \alpha/(D -4)$. As a result the above theory is free from Ostrogradski instability and a novel 4$D$ EGB gravity can be redefined
in the limit $D\to 4$ \cite{Glavan:2019inb}.  

\smallskip

Considering the limit $D\to 4$, for a static, spherically symmetric solution describing hydrostatic equilibrium of stars we choose the metric of the following form in Schwarzschild coordinates $t, r, \theta, \varphi$
\begin{eqnarray}\label{metric}
ds^2= - e^{2\Phi(r)}dt^2 + e^{2\Lambda(r)}dr^2 + r^{2}d\Omega_{2}^2,
\end{eqnarray} 
where $d\Omega_{2}^2$ is the metric of the unit 2-dimensional sphere. The metric functions $\Phi(r)$ and $\Lambda(r)$ depend on the radial coordinate $r$, respectively. Here we are interested in stars with anisotropic matter, the  energy-momentum tensor of which in $(3 + 1)$-dimensional space-times may be written down as follows 
\begin{equation}\label{emt}
T_{\mu\nu}=(\epsilon+P_{\perp})u_{\mu}u_{\nu}+ P_{\perp} g_{\mu\nu}+\left(P_r-P_{\perp}\right)\chi_{\mu}\chi_{\nu},
\end{equation}
where $u^{\mu}$ is the four-velocity and $\chi^{\nu}$ is the unit space-like vector in the radial direction. Here $\rho = \rho(r)$ is the energy density, $P = P(r)$ and $P_{\perp} = P_{\perp}(r)$ are the 
radial and transverse pressure, respectively.  Thus, in the limit $D \to 4$ the non-zero components of the field equations are the following
\begin{eqnarray}\label{DRE1}
&& \frac{2}{r} \frac{d\Lambda}{dr} = e^{2\Lambda} ~ \left[\frac{8\pi G}{c^4} \epsilon(r) - \frac{1-e^{-2\Lambda}}{r^2}\left(1-  \frac{\alpha(1-e^{-2\Lambda})}{r^2}\right)\right]\left[1 +  \frac{2\alpha(1-e^{-2\Lambda})}{r^2}\right]^{-1}, \\ 
&& \frac{2}{r} \frac{d\Phi}{dr} = e^{2\Lambda} ~\left[\frac{8\pi G}{c^4} P(r) + \frac{1-e^{-2\Lambda}}{r^2} \left(1- \frac{\alpha(1-e^{-2\Lambda})}{r^2} \right) \right] \left[1 +  \frac{2\alpha(1-e^{-2\Lambda})}{r^2}\right]^{-1},\label{DRE2} \\
&& \frac{dP}{dr} = - (\epsilon + P) \frac{d\Phi}{dr}+{2 \over r}\left(P_{\perp} - P\right).   \label{DRE3}
\end{eqnarray}

  To obtain the generalized TOV equations in a more familiar form, we replace the metric function with the following expression $e^{-2\Lambda}= 1-\frac{2G m(r)}{c^2 r}$, which represents the gravitational mass within the sphere of radius $r$. Now, using Eqs.  (\ref{DRE1}-\ref{DRE2}) and its derivative in equation (\ref{DRE3}), followed by some algebra, we obtain the following equation
\begin{equation}
{dP(r) \over dr} = -{G\epsilon(r) m(r) \over c^{2}r^2}\frac{\left[1+{P(r) \over \epsilon(r)}\right]\left[1+{4\pi r^3 P(r) \over c^{2}m(r)}-{2G\alpha m(r) \over c^{2}r^3}\right]}{\left[1+{4G\alpha m(r) \over c^{2}r^3}\right]\left[1-{2Gm(r) \over c^{2}r}\right]}+{2 \over r}\left(P_{\perp}(r) - P(r) \right) . \label{e2.11}
\end{equation}
One can recover the standard TOV equations of General Relativity for anisotropic fluid distribution when $\alpha \to 0$. Eq. (\ref{DRE1}) can be rewritten in terms of the mass function $m(r)$ as follows
\begin{equation}
m'(r)=\frac{6  \alpha G m(r)^2+4 \pi  r^6 \epsilon (r)}{4 \alpha G r m(r)+c^2 r^4}, \label{e2.12}
\end{equation}
where prime denotes differentiation with respect to radial coordinate.
Here, we impose $m(r=0)=0$ to be the appropriate condition at the centre of the fluid sphere.  Further, it is convenient to work with dimensionless variables, and thus we introduce the following set of dimensionless parameters: $P(r)=\epsilon_{0}{\bar P}(r)$ and $\epsilon(r)=\epsilon_{0}{\bar \epsilon}(r)$ and $m(r)=M_{\odot}{\bar M}(r)$, with $\epsilon_{0}=1\,{\rm MeV}/{\rm fm}^{3}$. After a short calculation, the above two equations take now the form
\begin{eqnarray}
{d{\bar P}(r) \over dr} = - \frac{c_1 {\bar \epsilon}(r) {\bar M}(r)}{r^2} \frac{\left[1+{{\bar P}(r) \over {\bar \epsilon}(r)}\right] \left[1+{c_2 r^3 {\bar P}(r) \over {\bar M}(r)}-{2 c_1 \alpha {\bar M}(r) \over r^3}\right] }{\left[1+{4 c_1 \alpha {\bar M}(r) \over r^3}\right]\left[1-{2 c_1 {\bar M}(r) \over r}\right]}+{2 \over r}\left( \bar P_{\perp} (r) - \bar P(r) \right) , \label{e2.11d}
\end{eqnarray}
and
\begin{eqnarray}
\frac{d{\bar M}(r)}{dr} = \frac{6 c_1 \alpha {\bar M}(r)^2 + c_2 r^6 {\bar \epsilon}(r)}{4 c_1 \alpha r {\bar M}(r) + r^4}, \label{mr}
\end{eqnarray}
where $c_1 \equiv \frac{G M_{\odot}}{c^2} = 1.474 \text{ km}$ and $c_2 \equiv \frac{4 \pi \epsilon_0}{M_{\odot} c^2} = 1.125 \times 10^{-5} \; \text{km}^{-3}$. Those equations can be solved numerically for a given EoS $P = P(\epsilon)$ that relates the pressure with the energy density. Next, we review the structure equations describing hydrostatic equilibrium of a QCD motivated EoS, which allow us to obtain quark star interior solutions.

\section{Equations of State for Quark matter } \label{sec3}

The predicted masses of pulsars through the observations put a strong constraint on the NS matter EoS, and consequently on the composition and inner structure of NSs. Thus, mass-radius profiles and the maximum mass of NSs are crucial in our effort to understand the true nature of matter inside a star, and also set the limit from where the formation of a black hole is unavoidable. Recent instrumentation and computational advances speculate that when nuclear matter is compressed at ultrahigh densities and temperatures, the nucleon cores substantially overlap. Due to a quark nova, i.e. an explosive transition, a more stable configuration made of de-confined quarks is expected to emerge, rather than a NS made of hadrons and electrons. At such an extreme situation one would expect that most NSs consist of conventional quark matter cores. This observation was pointed out by Ivanenko \& Kurdgelaidze \cite{Ivanenko} in the 60's as a possible existence of a \textit{quarkian core} in very massive stars.

\smallskip

Even more intriguing is the existence of quark matter in a NS, is the possible existence of a new, less conventional class of compact stars, the so called \textit{strange stars}. In the original model it was believed that dense matter in the core of compact stars consists of three fundamental quarks nicknamed up $(u)$ down $(d)$ and strange $(s)$, and a small number of electrons to ensure charge neutrality satisfying the Bodmer-Witten hypothesis \cite{Haensel:1986qb,Alcock:1986qb}. From then on, QSs are treated as a kind of compact objects composed of pure (strange) quark matter, and they should be described viewing color quarks as the fundamental degree of freedom. In earlier investigations, the MIT Bag Model is one of the most successful phenomenological models that characterizes a degenerated Fermi gas of quarks up, down and strange \cite{Chodos1974,Farhi1984}, and bag constant $B$ actually the pressure difference between the two phases. The simplicity of this model makes it attractive and more frequently used in QSs studies.

\smallskip

Nevertheless, recent developments in dense matter physics led to the construction of several models based on Quantum Chromodynamics (QCD), suggesting that quark matter might be in different color superconducting phases \cite{Alford:2002kj,Alford:2005wj,Mannarelli:2007bs}.
Thus, QCD corrections of second and fourth order require an approximate characterization of confined quarks 
\cite{Flores:2017kte}. Following the EoS discussed in \cite{Flores:2017kte}, we present the EoS that consists of homogeneous and unpaired, overall electrically neutral, 3-flavor interacting quark matter. But for simplicity, we describe this phase using the simple thermodynamic Bag model EoS \cite{Alford:2004pf} with $\mathcal{O}$ $(m_s^4)$ corrections. In previous works several authors explored QSs stars made up of interacting quark EoS at ultra-high densities have also been suggested and intensively investigated \cite{Becerra-Vergara:2019uzm,Banerjee:2020dad}. Finally, the interacting quark EoS reads \cite{Becerra-Vergara:2019uzm}
\begin{equation} \label{Prad1}
P=\dfrac{1}{3}\left(\epsilon-4B\right)-\dfrac{m_{s}^{2}}{3\pi}\sqrt{\dfrac{\epsilon-B}{a_4}}
+\dfrac{m_{s}^{4}}{12\pi^{2}}\left[1-\dfrac{1}{a_4}+3\ln\left(\dfrac{8\pi}{3m_{s}^{2}}\sqrt{\dfrac{\epsilon-B}{a_4}}\right)\right],
\end{equation}
where $\epsilon$ is the energy density of homogeneously distributed quark matter  (also to $\mathcal{O}$ $(m_s^4)$ in the Bag model). In the present work, we have employed the values of Bag constant $B$ within the range  $57\leq B \leq 92$ MeV/fm$^3$ \cite{Burgio:2018mcr,Blaschke:2018mqw}, and the quark strange mass is  $m_{s}$ to be $100 \,{\rm MeV}$ \cite{Beringer:2012}. Finally, $a_4$  is the parameter that comes from the QCD corrections on the pressure of the quark-free Fermi sea; directly related with the  $M-R$ relation of QSs.

\smallskip

As we are considering anisotropic stars, we propose a new generalized EoS for the tangential pressure followed by \cite{Becerra-Vergara:2019uzm}:
\begin{eqnarray} \label{Prad2}
P_{\perp}= P_c +\dfrac{1}{3}\left(\epsilon-4B_{\perp}\right)-\dfrac{m_{s}^{2}}{3\pi}\sqrt{\dfrac{\epsilon-B_{\perp}}{a_4^{\perp}}}
+\dfrac{m_{s}^{4}}{12\pi^{2}}\left[1-\dfrac{1}{a_4^{\perp}}+3\ln\left(\dfrac{8\pi}{3m_{s}^{2}}\sqrt{\dfrac{\epsilon-B_{\perp}}{a_4^{\perp}}}\right)\right] \nonumber\\
-\dfrac{1}{3}\left(\epsilon_c-4B_{\perp}\right)+\dfrac{m_{s}^{2}}{3\pi}\sqrt{\dfrac{\epsilon_c-B_{\perp}}{a_4^{\perp}}}
-\dfrac{m_{s}^{4}}{12\pi^{2}}\left[1-\dfrac{1}{a_4^{\perp}}+3\ln\left(\dfrac{8\pi}{3m_{s}^{2}}\sqrt{\dfrac{\epsilon_c-B_{\perp}}{a_4^{\perp}}}\right)\right],
\end{eqnarray}
where $\epsilon_c$ and $P_c$ are the central energy density and central radial pressure for Eq. (\ref{Prad1}), respectively. Note that at the center of the star i.e., $r=0$ the  radial and tangential pressures are the same for $B = B_{\perp}$ and  $a_4 =a_4^{\perp}$. This condition represents the case of an isotropic fluid and desirable for any anisotropic fluid sphere. 
Meanwhile, in the rest of the star, if the tangential and radial pressure are equal, the expression  corresponding to the pressure will have contributions only from the first term. The domain space of the 
parameters $a_4^{\perp}$ and $B_{\perp}$ associated with the expression (\ref{Prad2}) lie in the  same range of $a_4$ and $B$.

\section{Numerical results}\label{sec4}

To solve the generalized TOV equations numerically for $M(r)$ and $P(r)$ one can integrate outwards starting from the origin $(r = 0)$ to the surface of the stars, $r = R$, where the pressure vanishes. To that end, we solve the
equations  (\ref{e2.11d}) and (\ref{mr}) employing the EoSs (\ref{Prad1}) and (\ref{Prad2}) to  calculate the maximum mass and other properties of the QSs. Those equations must be solved in view of the initial conditions $P(r_{0}) = P_{c}$ and $M(r_{0}) = 0$, where $P_{c}$ is the central pressure. To avoid the discontinuities that appear in the denominators, we set  $r_{0}= 10^{-5}$ and mass $m(r_{0})= 10^{-30}$ rather than zero. Additionally, we adopt the unit conversion $1\,{\rm fm^{-1}}  = 197.3\,{\rm MeV}$ to synchronize each term given in Eqs. (\ref{Prad1})
and (\ref{Prad2}), respectively.

\smallskip

Now we report the results of our computations. We have performed a numerical work for several values of $a_4^{\perp}$ and $B_{\perp}$ i.e., we pay special attention on the parameter $a_4^{\perp}$ ranging from $a_4^{\perp} = 0.1$ to $a_4^{\perp} = 0.9$, which is related to the maximum mass of the star around $2M_{\odot}$ at $a_4\approx 0.7$, as suggested in \cite{Fraga:2001id}. For the entire computation we consider a certain value of central pressure,  $P({r_{0}})=700 \,{\rm MeV}/{\rm fm}^{3}$, and the radius of the star is identified when the pressure vanishes or drops to a very small value. Besides that, we vary the GB constant $\alpha$, which is taken to take values in the interval $-5\,{\rm km}^{2} \leq \alpha \leq 5\,{\rm km}^{2}$ following \cite{Banerjee:2020dad}. Our results show that the gravitational mass of star grows with $\alpha$ within the $4D$ EGB model.

\section{Mass-radius relation}\label{sec5}

In this section we present and discuss our main numerical results summarized in the figures below. We investigate in detail the impact of the free parameters of the theory on properties of anisotropic SQSs, such as mass, radius, factor of compactness and binding energy. The free parameters in this work are on the one hand the Gauss-Bonnet coupling (from the gravity side), and on the other hand $B, a_4, B^{\perp}, a_4^{\perp}$ that enter into the EoS of quark matter. The mass of the stars is measured in solar masses, the radius in $\textrm{km}$, while the energy density, the pressure and the bag constant are measured in $MeV/fm^3$. Finally, $a_4, a_4^{\perp}$ are dimensionless, and the GB coupling is measured in $km^2$.

\begin{figure}[h]
    \centering
    \includegraphics[width = 8.3 cm]{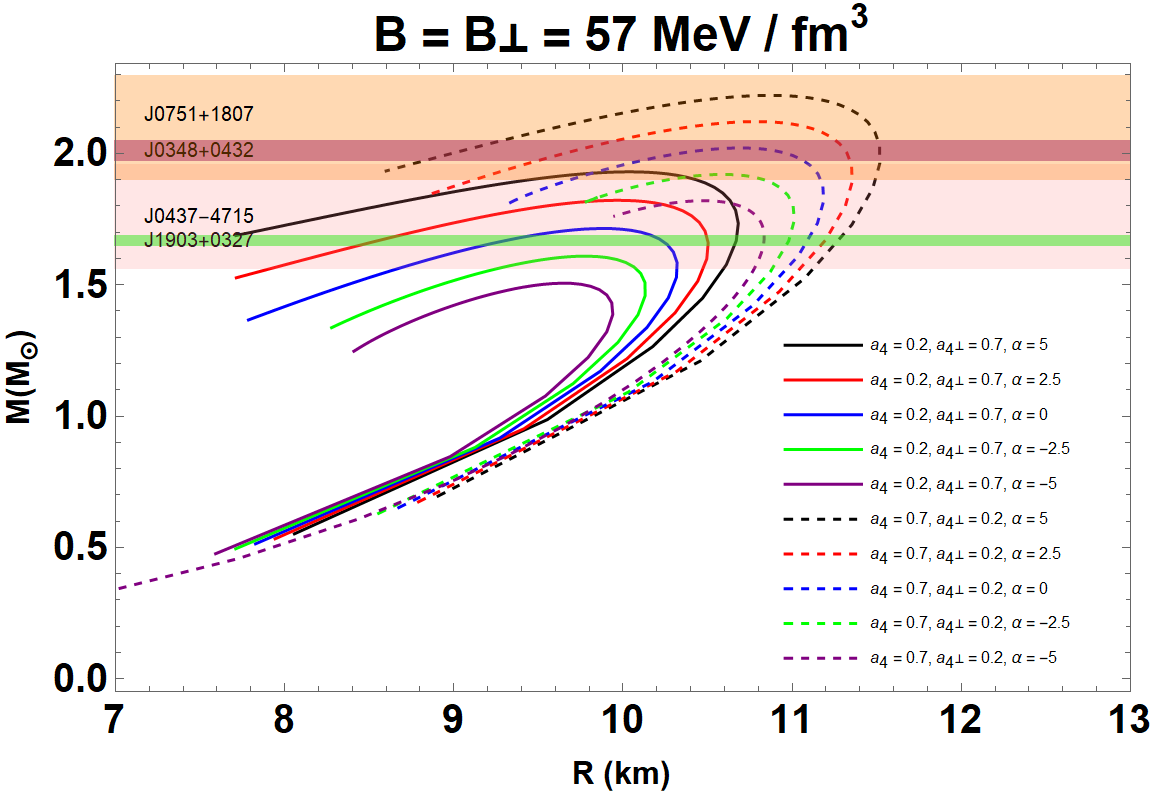}
    \includegraphics[width = 8.3 cm]{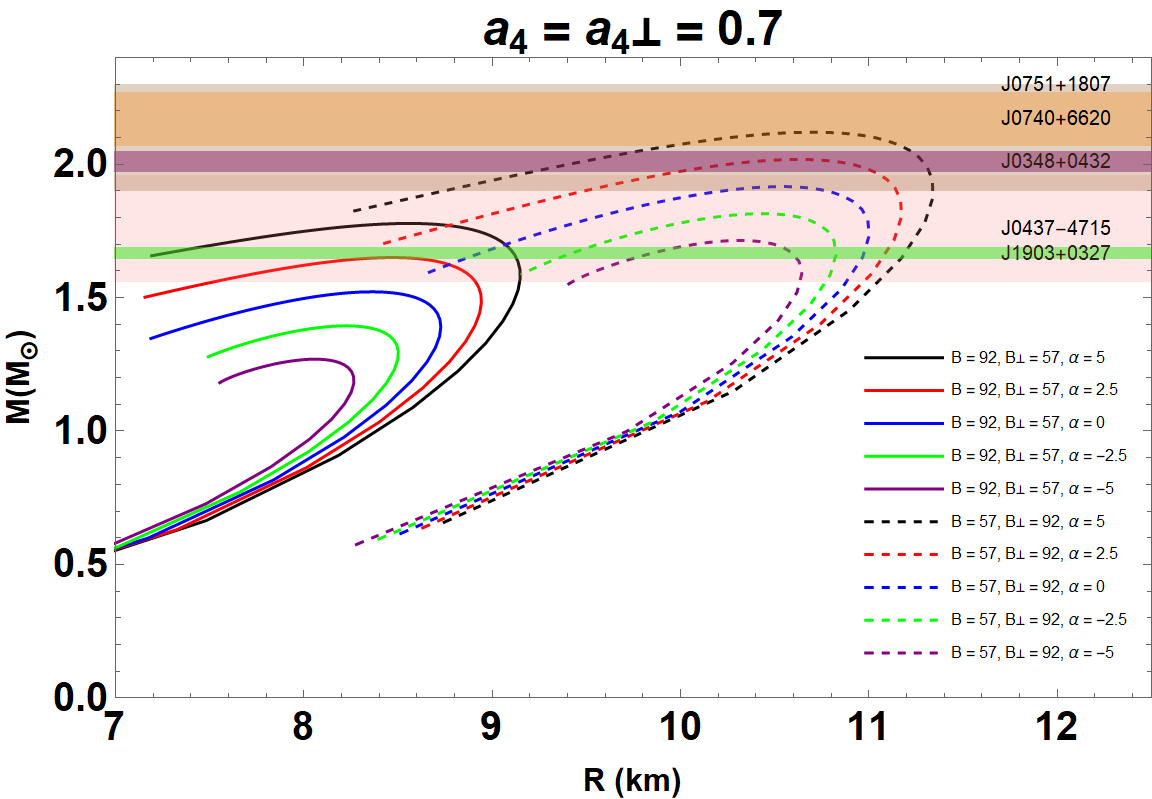}
    \caption{Mass-radius relation for anisotropic quark stars, for (i) $B = B_{\perp}$ when $ a_4 \neq a_4^{\perp}$ in the left panel and (ii) $B \neq B_{\perp}$ when $ a_4 = a_4^{\perp}$ in the right panel, respectively. The horizontal bands show the observational constraints from various pulsar measurements: PSR J0348+0432 (Purple) \cite{Antoniadis:2013pzd},  PSR J0751+1807 (Orange) \cite{Nice:2005fi}, PSR J0437-4715 (Pink) \cite{Verbiest:2008gy} and  PSR J1903+0327 (Green) \cite{Freire2011}.}
    \label{Total_MR_fix_B}
\end{figure}
\smallskip

First, we show the $M-R$ relationships, in which we also include the masses of the following observed pulsars: PSR J0348+0432 with $2.01 \pm 0.04 M_{\odot}$ (Purple) \cite{Antoniadis:2013pzd},  PSR J0751+1807 with $2.1 \pm 0.2 M_{\odot}$ (Orange) \cite{Nice:2005fi}, PSR J0437-4715 with $1.76 \pm 0.20 M_{\odot}$ (Pink) \cite{Verbiest:2008gy} and  PSR J1903+0327 with $1.667\pm 0.021\, M_{\odot}$ (Green) \cite{Freire2011}. 

\begin{enumerate}
    \item In the left panel of Fig.~\ref{Total_MR_fix_B} we fix the bag constants, and vary $\alpha$ and $a_4, a_4^{\perp}$ setting $B = 57~MeV/fm^3=B_{\perp}$. The first group of profiles (solid curves) corresponds to $a_4 = 0.2, a_4^{\perp} = 0.7$, and progressively increasing values of the Gauss-Bonnet coupling, both negative and positive, from $\alpha = -5~km^2$ to $\alpha = 5~km^2$. For comparison reasons the GR case ($\alpha = 0$) is shown as well. The profiles exhibit the usual form expected for quark matter with a maximum radius and a maximum mass shown in Table I for $\alpha = 5~km^2$. We report a maximum mass $M_{max} = 1.93~M_{\odot}$ and a corresponding radius $R_* = 10.04~km$. Both the maximum radius and the maximum mass of the stars grow with $\alpha$. 
    \item The second group of profiles (dashed curves) in the same panel of Fig.~\ref{Total_MR_fix_B}, corresponds to $a_4 = 0.7, a_4^{\perp} = 0.2$, and progressively increasing values of the Gauss-Bonnet coupling from $\alpha = -5~km^2$ to $\alpha = 5~km^2$. Similarly to the first group, the profiles exhibit the usual form expected for quark matter with a maximum radius and a maximum mass that grow with $\alpha$. Only the second group crosses the two solar mass bound set by the massive pulsar J0348+0432, however we observe a good agreement with other pulsars detected, such as PSR J0437-4715 and PSR J1903+0327.
    \item Next, we move to the right panel of Fig.~\ref{Total_MR_fix_B}, where we fix $a_4, a_4^{\perp}$, and we vary $\alpha$ and $B, B_{\perp}$ setting $a_4 = 0.7 = a_4^{\perp}$. 
The first group of profiles (solid curves) corresponds to $B = 92~MeV/fm^3, B^{\perp} = 57~MeV/fm^3$, and progressively increasing values of the Gauss-Bonnet coupling, both negative and positive, from $\alpha = -5~km^2$ to $\alpha = 5~km^2$. For comparison reasons the GR case ($\alpha=0$) is shown as well. Similarly to the left panel, the profiles exhibit the usual form expected for quark matter with a maximum radius and a maximum mass, which grow with $\alpha$.
    \item Finally, the second group of profiles (dashed curves) corresponds to $B = 57~MeV/fm^3, B_{\perp} = 92~MeV/fm^3$, and progressively increasing values of the Gauss-Bonnet coupling, from $\alpha = -5~km^2$ to $\alpha = 5~km^2$. Once again, we observe the same features similarly to the left panel, namely, i) the maximum mass and radius grow with $\alpha$, and ii) although only the second group crosses the two solar masses bound, we observe a good agreement with other pulsars detected, such as PSR J0437-4715 and PSR J1903+0327.
\end{enumerate}

\begin{figure}[h]
    \centering
    \includegraphics[width = 8 cm]{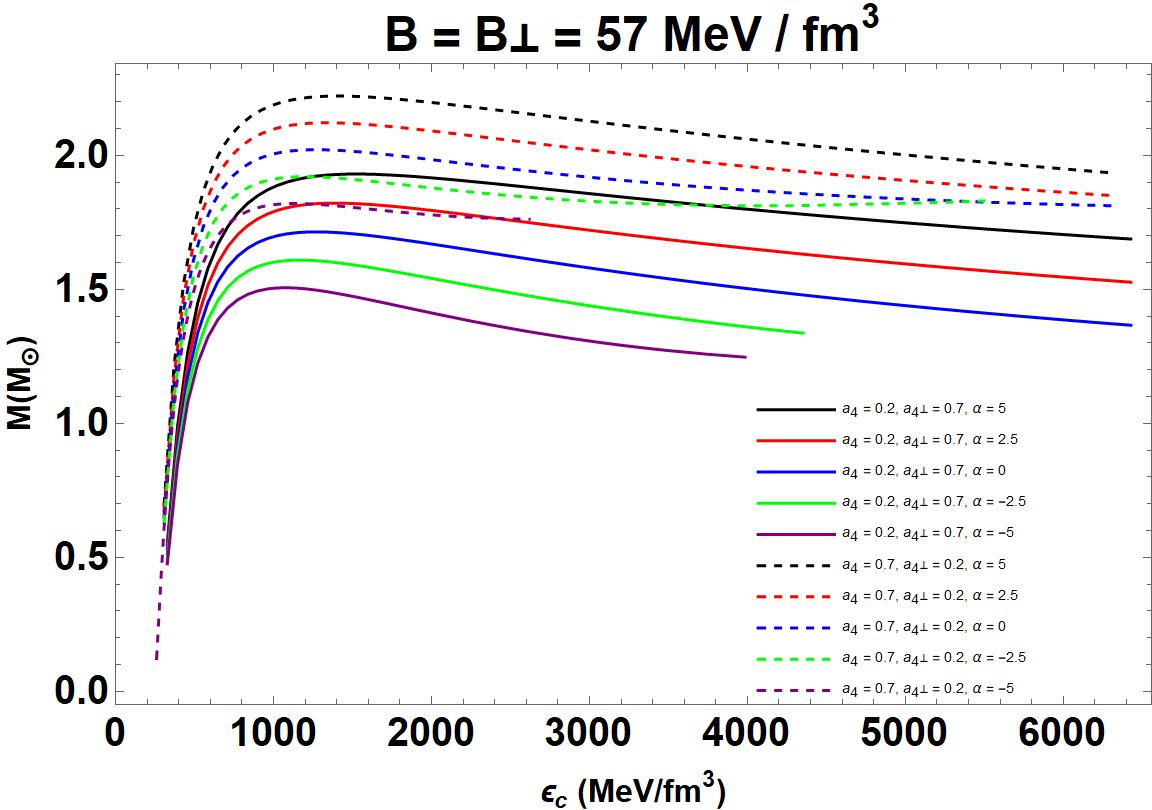}
    \includegraphics[width = 8 cm]{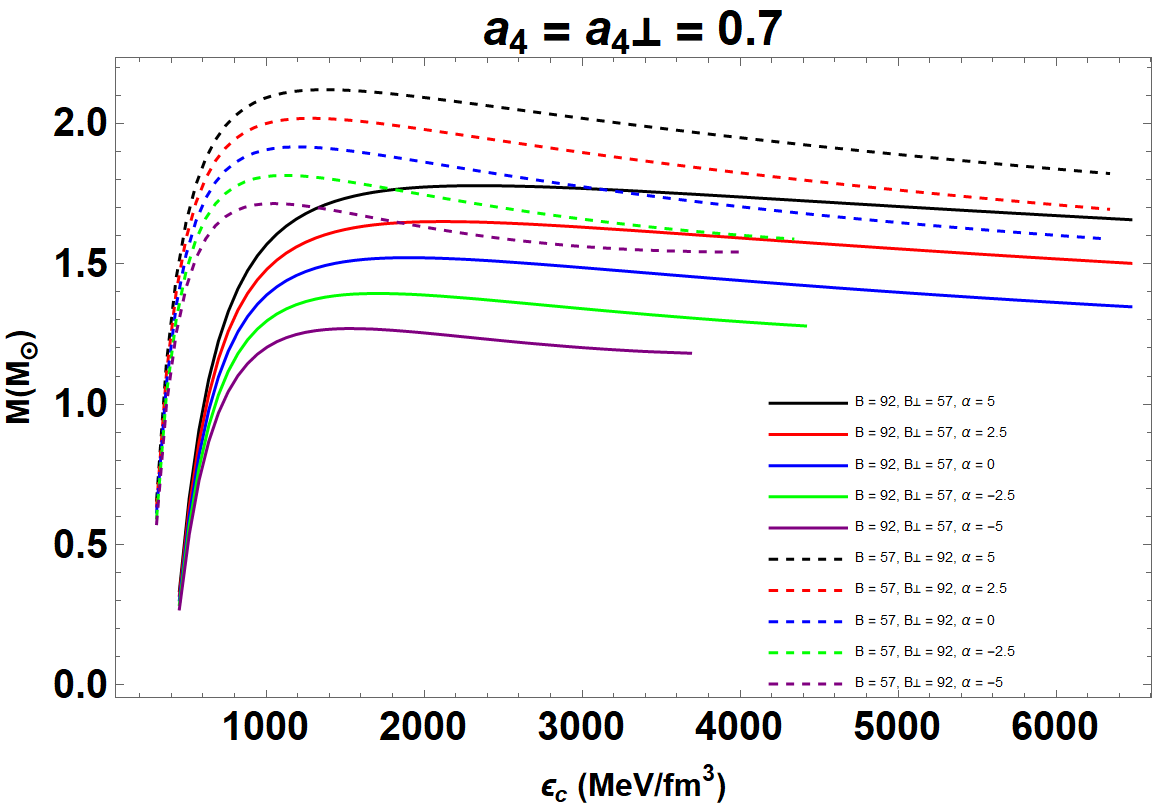}
    \caption{Variation of the total mass versus the central energy density for different values of the coupling constant $\alpha$. }
    \label{Total_ME_fix_B}
\end{figure}

\begin{figure}[h]
    \centering
    \includegraphics[width = 8 cm]{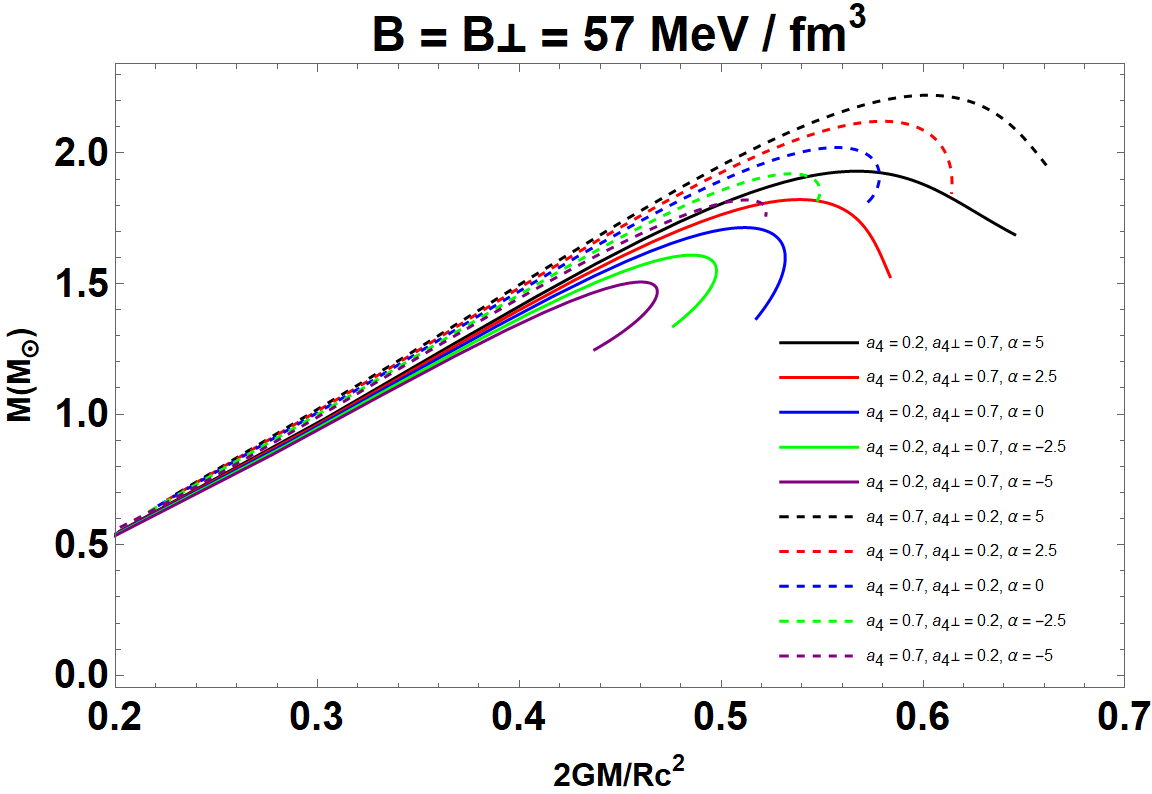}
    \includegraphics[width = 8 cm]{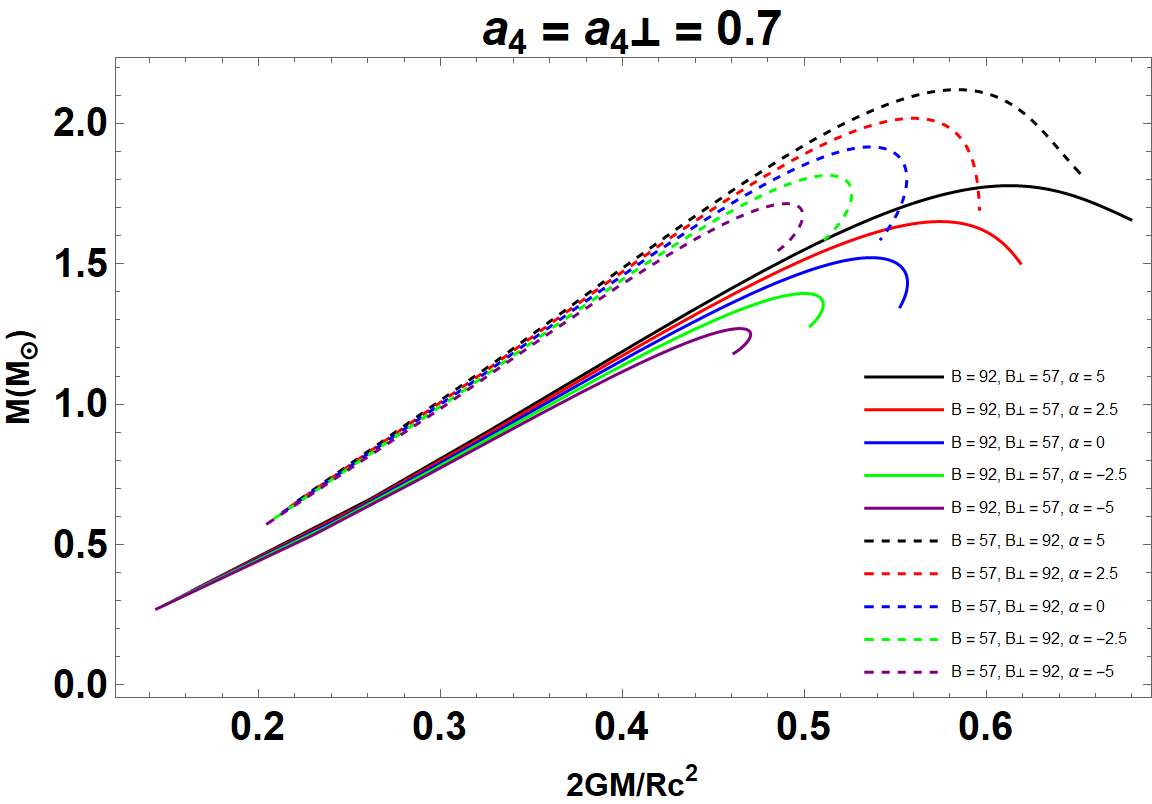}
    \caption{Variation of the maximum mass versus the $2GM/c^2 R$ (compactness). }
    \label{Total_Com_fix_B}
\end{figure}

\smallskip

\section{Properties of stars other than $M-R$ profiles}\label{sec6}

Next, in Figs.~\ref{Total_ME_fix_B}  and \ref{Total_Com_fix_B} we show the mass of the stars versus central energy density, and the mass of the stars versus the factor of compactness, $C = 2GM / (R c^2)$, respectively. In each figure there are two panels showing the same two groups of curves corresponding to the same values of the parameters, precisely as in Fig.~\ref{Total_MR_fix_B}. 

\smallskip

To plot Fig. ~\ref{Total_ME_fix_B} we take central energy density to be in the range 
$(500-6500)~MeV/fm^3$. The mass of the star reaches a maximum value at $\epsilon_c^*$, which grows with the GB coupling, in agreement with Fig.~\ref{Total_MR_fix_B}. Moreover, according to the  Harrison-Zeldovich-Novikov criterion \cite{harrison,ZN}
\begin{equation}
\frac{dM}{d \rho_c} > 0  \; \; \; \rightarrow \textrm{stable configuration}
\end{equation}
\begin{equation}
\frac{dM}{d \rho_c} < 0  \; \; \; \rightarrow \textrm{unstable configuration}
\end{equation}
only the first part of the curve, before the maximum value, corresponds to a stable configuration. Therefore, the point at the extremum of the curve separates the stable from the unstable configuration.

\smallskip

What is more, according to Fig.~\ref{Total_Com_fix_B}, the factor of compactness, $C$, reaches higher values as $\alpha$ increases, while the Buchdahl bound, $C \leq 8/9$ \cite{buchdahl}, is not violated. The factor of compactness allows us to compute the surface red-shift, $z_s$, which is defined by
\begin{equation}
z_s = -1 + (1-C)^{-1/2},
\end{equation}
and which is of great importance to astronomers, since it is related to the emission produced by photons from the surface of the star. Some values of $C$ as well as the corresponding values of $z_s$ are shown in Table \ref{table1}. In particular, the limits on the coupling $\alpha$ obtained in \cite{Clifton:2020xhc} are more 
stringent than those obtained in \cite{Feng:2020duo}, so let us elaborate on those.  The sources used are shown in Table \ref{table1}, and as expected some put stronger bounds than others.  Overall, the authors have found a conservative as well as a less conservative bound,  $|\alpha| \leq 10^{10}~m^2$ or $|\alpha| \leq 10^{8}~m^2$. In our work,  the numerical values of $\alpha$ we have considered are compatible with those limits. 

\smallskip

\begin{figure}[h]
    \centering
    \includegraphics[width = 7.5 cm]{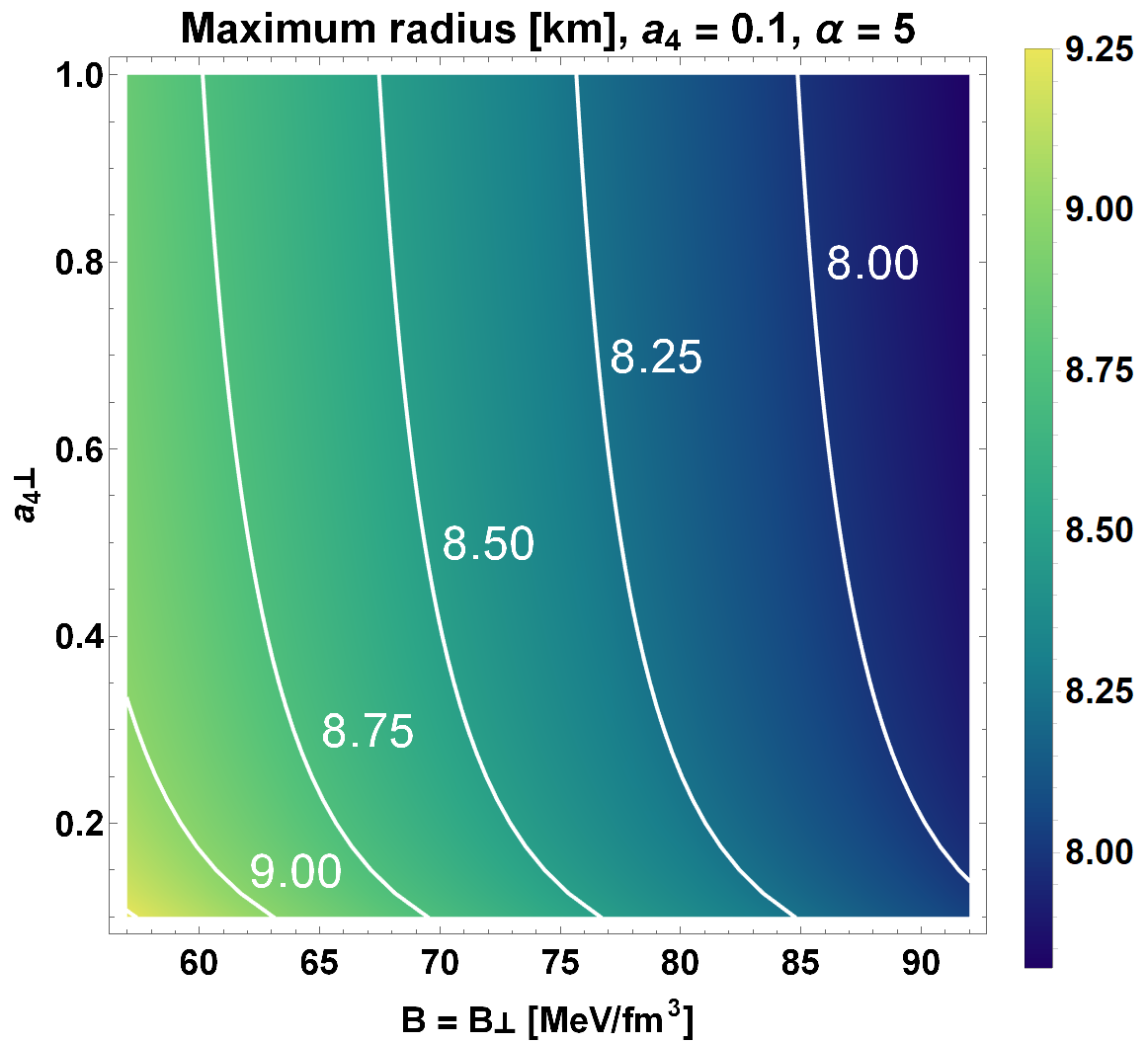}
    \includegraphics[width = 7.5 cm]{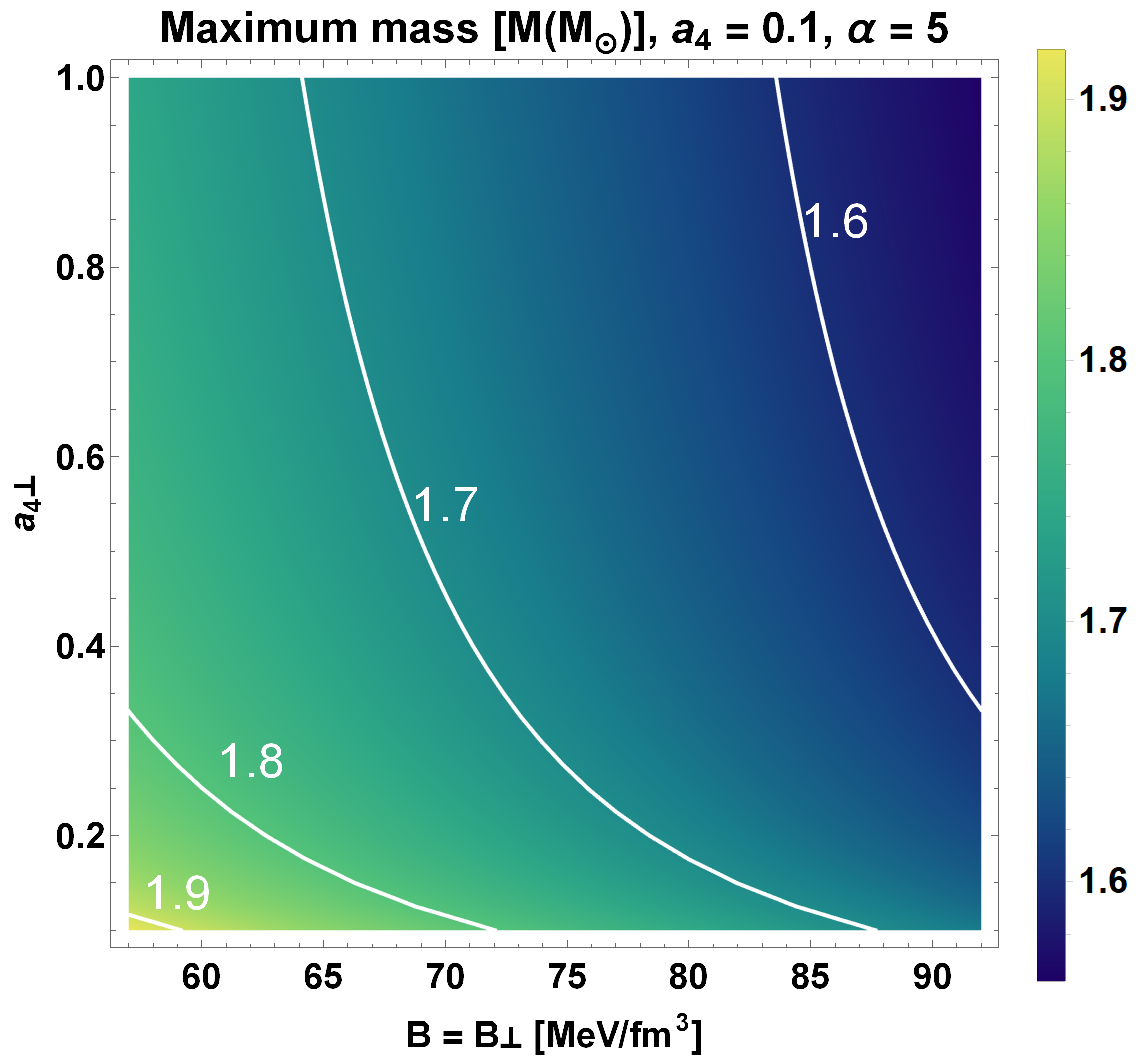}
    \includegraphics[width = 7.5 cm]{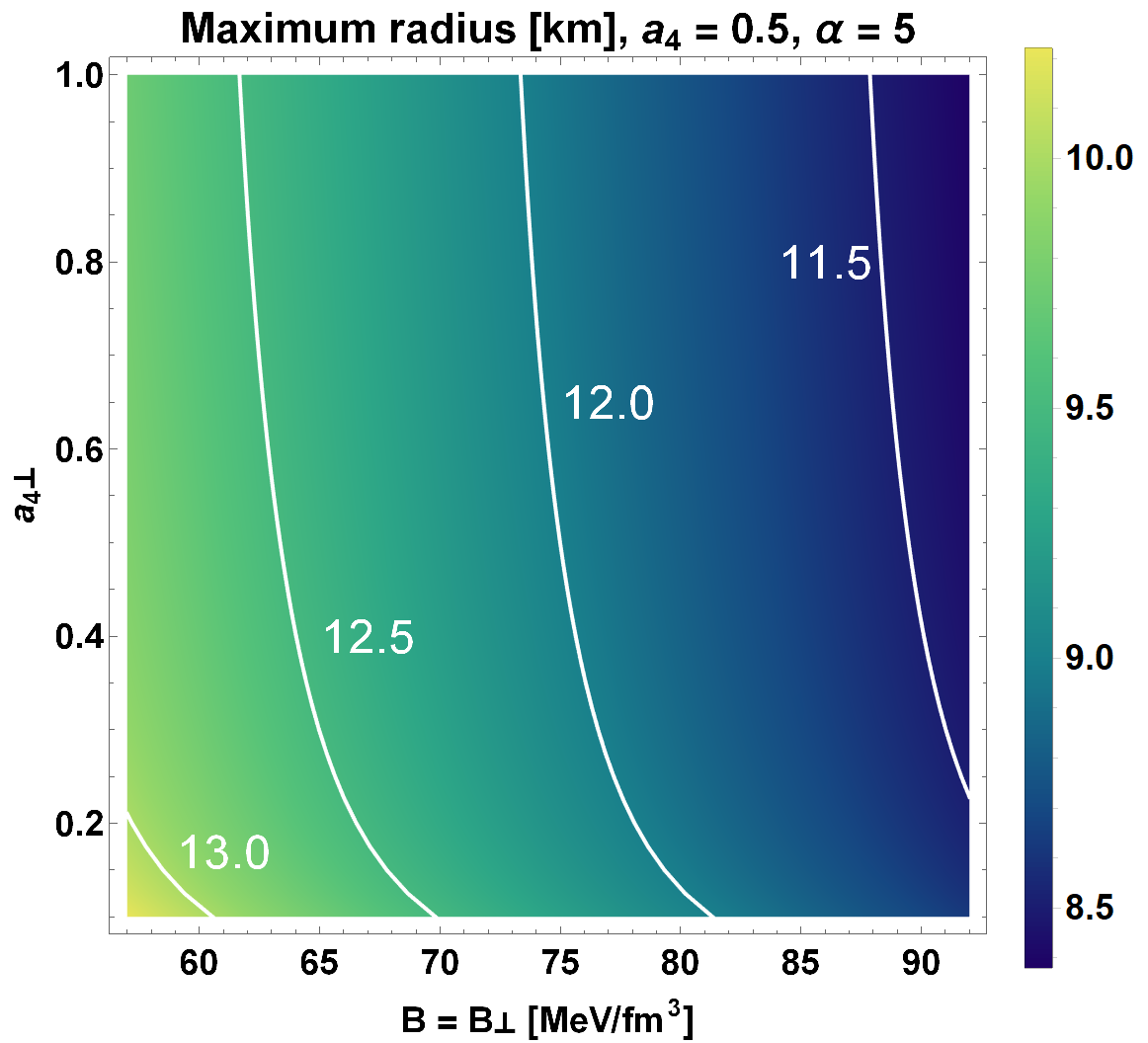}
    \includegraphics[width = 7.5 cm]{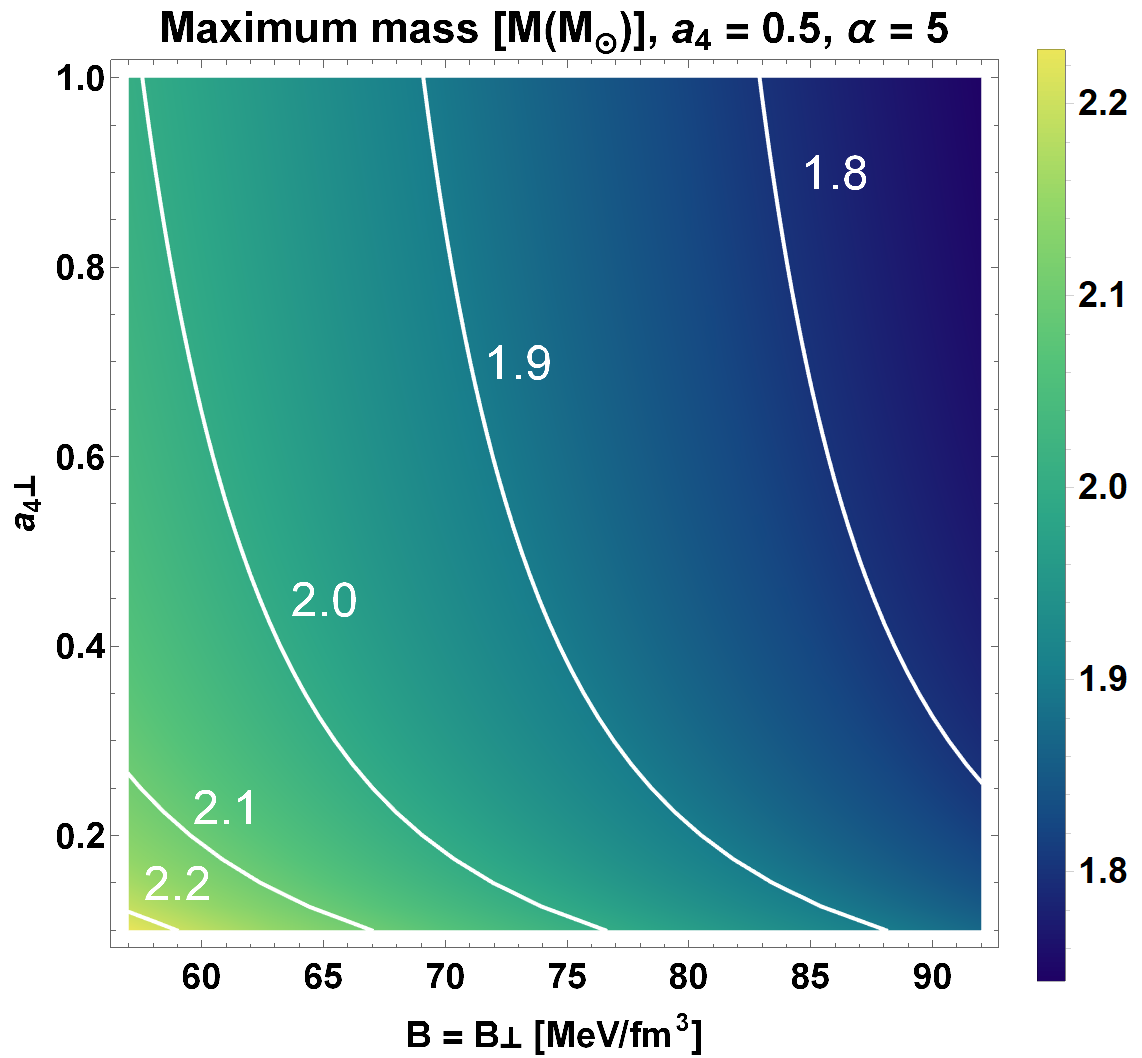}
    \includegraphics[width = 7.5 cm]{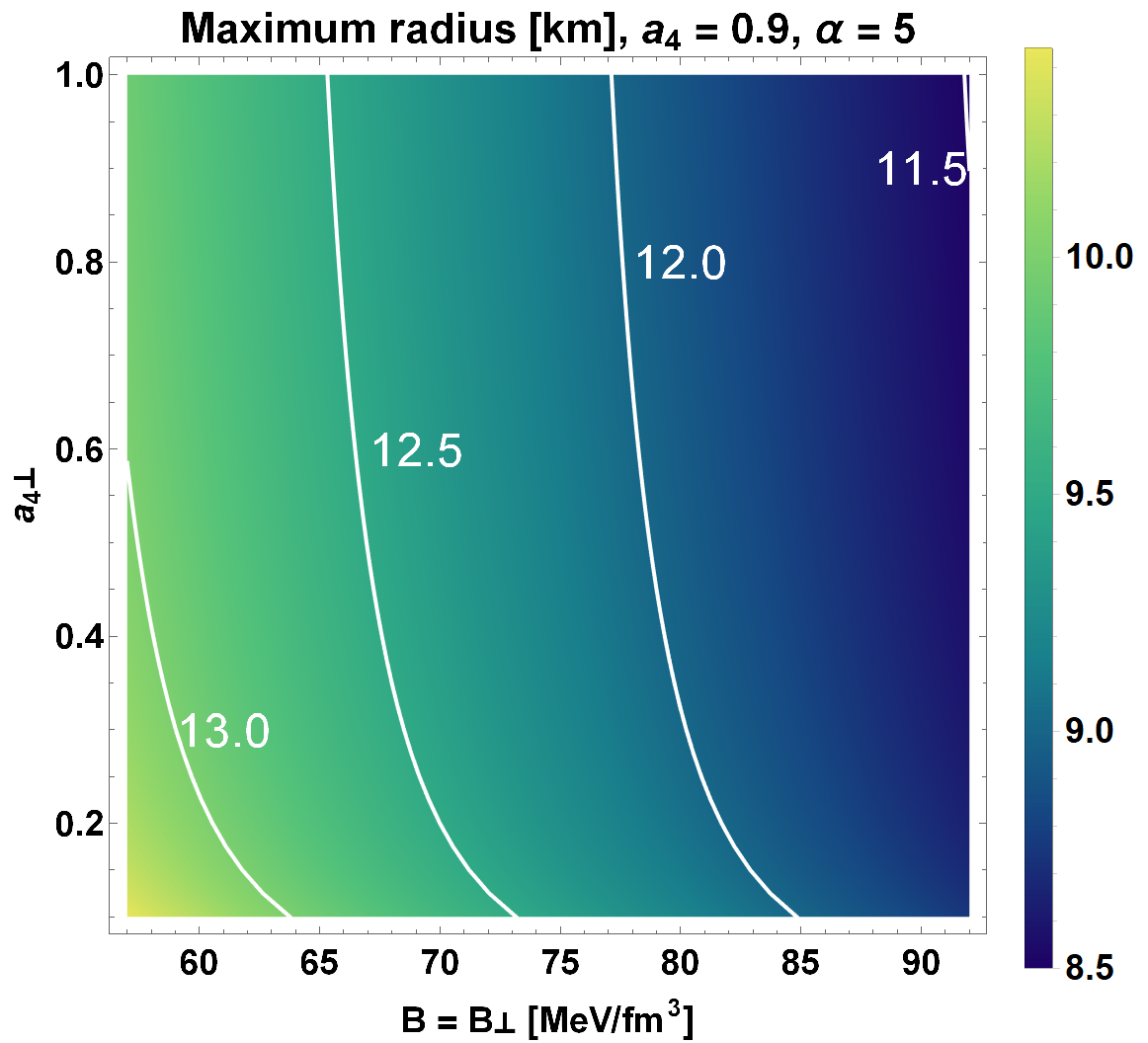}
    \includegraphics[width = 7.5 cm]{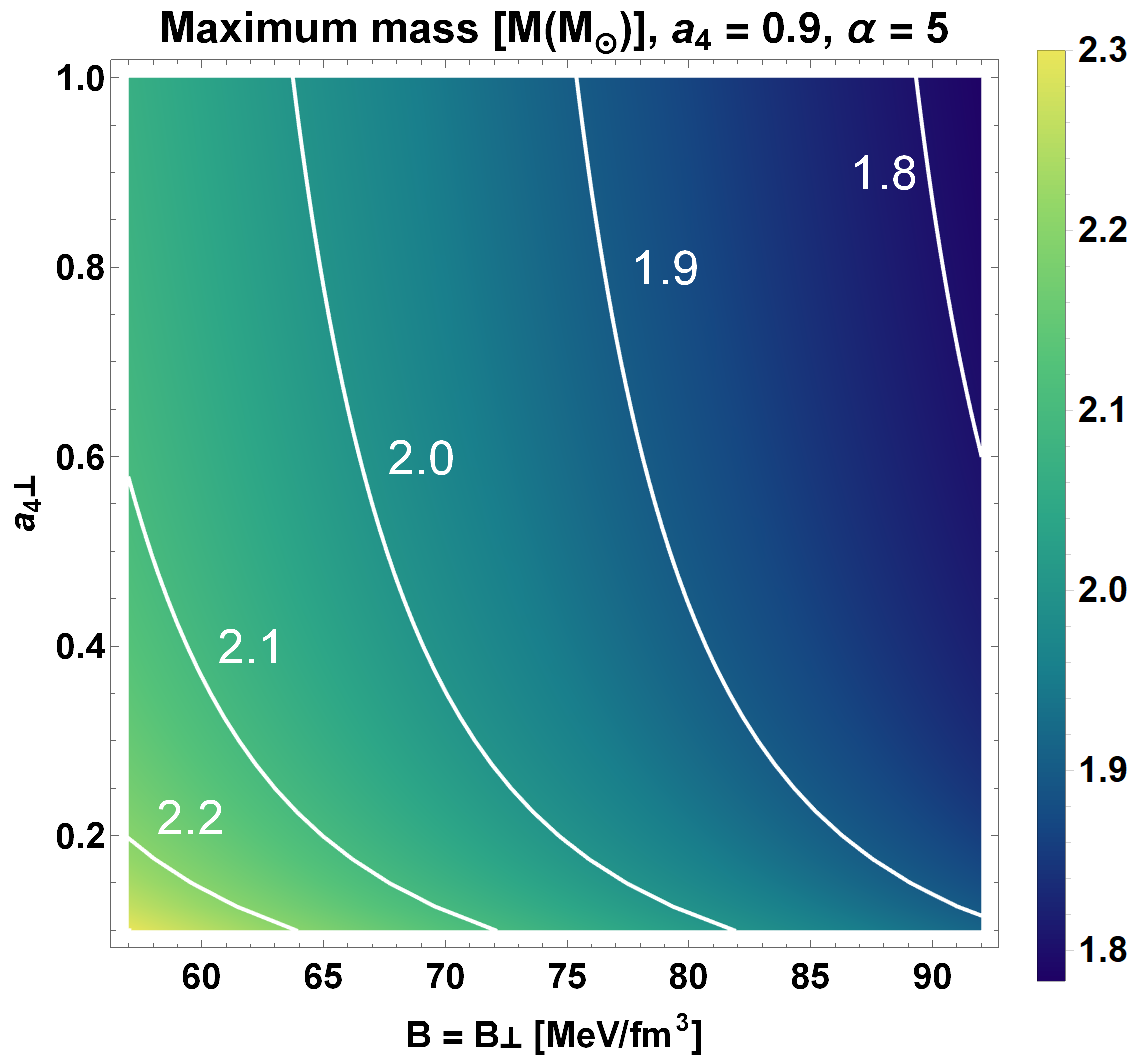}
    \caption{Maximum masses and their corresponding radii for the full range of values of $B_{\perp}$ and $ a_4^{\perp}$. The contour plot for $57 < B = B_{\perp} < 92$ with varying $ a_4^{\perp}$ in the range of $0< a_4^{\perp} < 1$  for $a_{4} = 0.1$, $0.5$ and $0.9$, respectively. We recorded for $\alpha = 5$ to get maximum effect on mass radius relation. }
    \label{contour}
\end{figure}

Moreover, in Fig.~\ref{contour} we show the maximum radius in $km$ (left column) as well as the maximum mass in solar masses (right column) of the stars on the $B-a_4^{\perp}$ plane setting $\alpha = 5~km^2$ and $a_4 = 0.1$ (first row), $a_4 = 0.5$ (second raw) and $a_4 = 0.9$ (third row). The constraint $M > 2 M_{\odot}$ is satisfied for $B < 76~MeV/fm^3$ when $a_4 = 0.5$ and for $B < 82~MeV/fm^3$ when $a_4 = 0.9$. 

\smallskip

\begin{figure}[h]
    \centering
    \includegraphics[width = 7.5 cm]{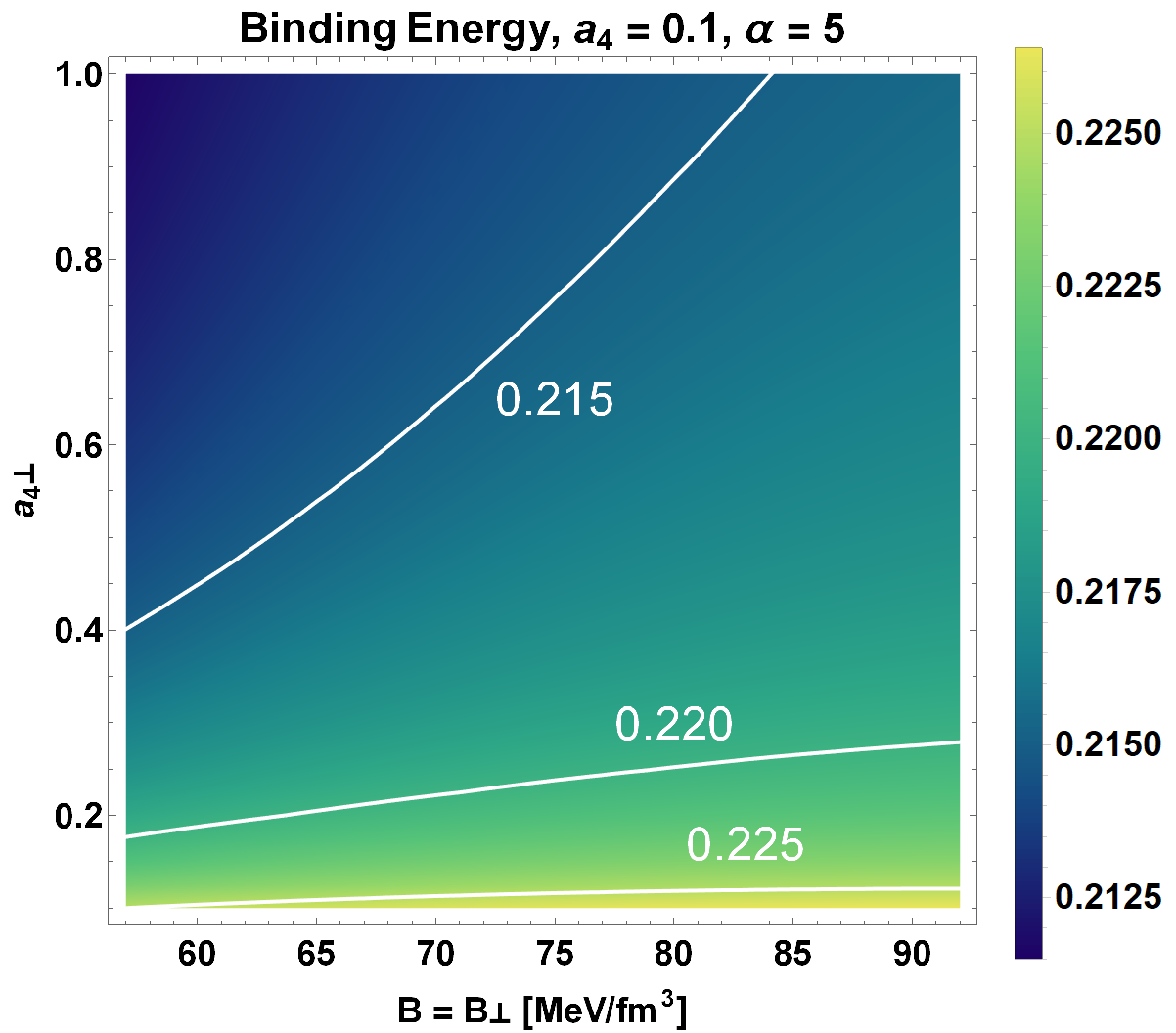}
    \includegraphics[width = 7.5 cm]{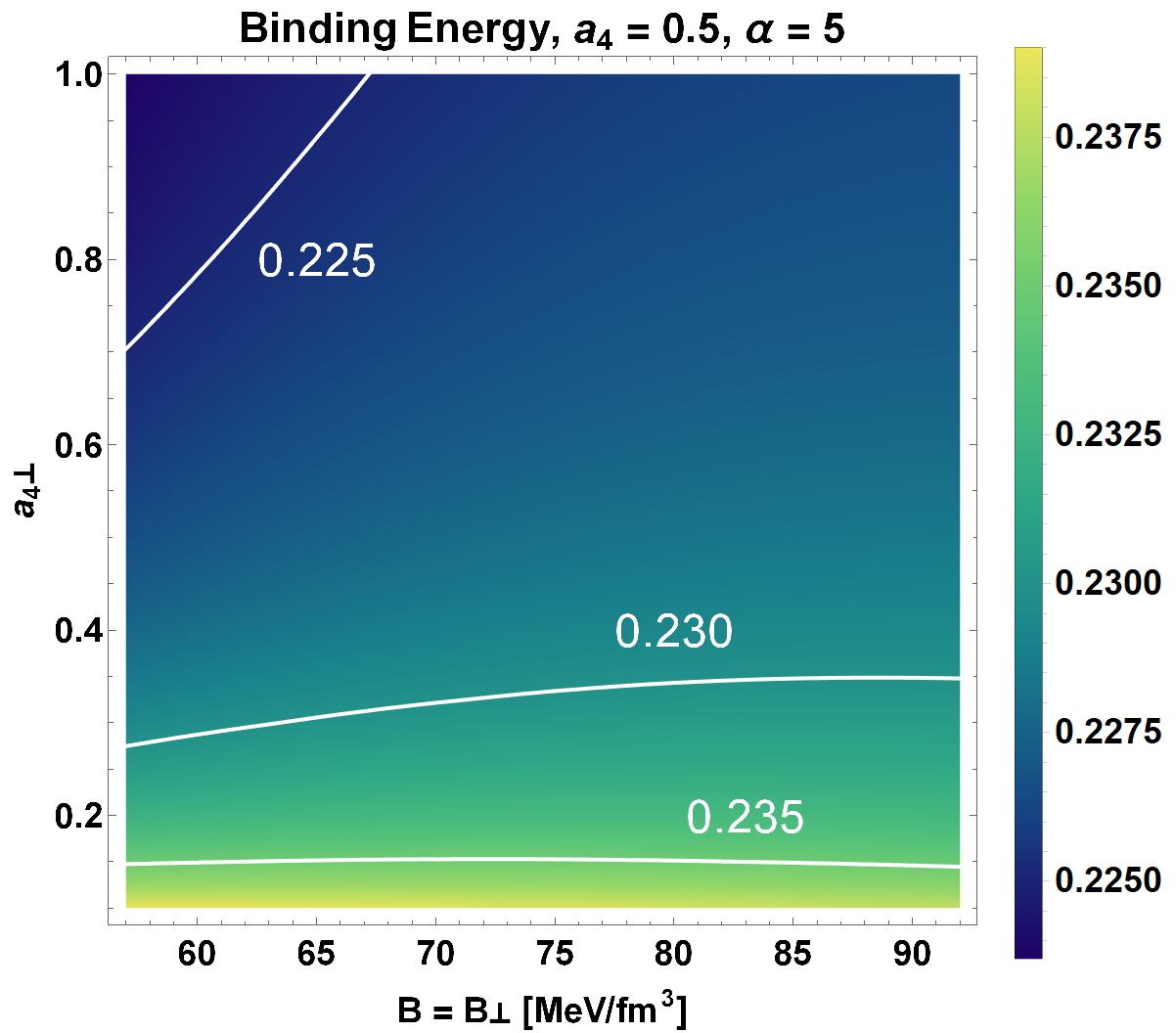}
    \includegraphics[width = 7.5 cm]{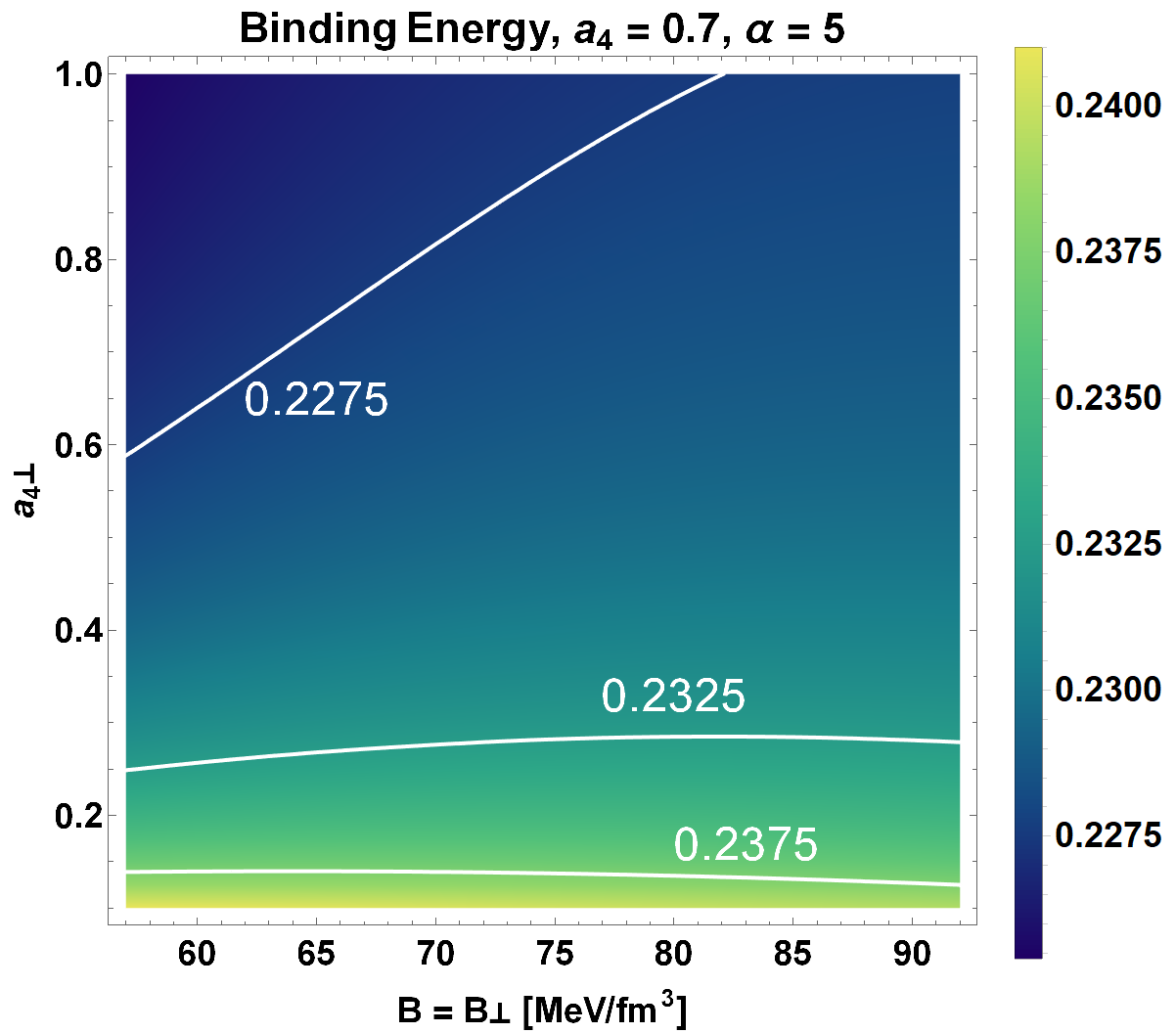}
    \includegraphics[width = 7.5 cm]{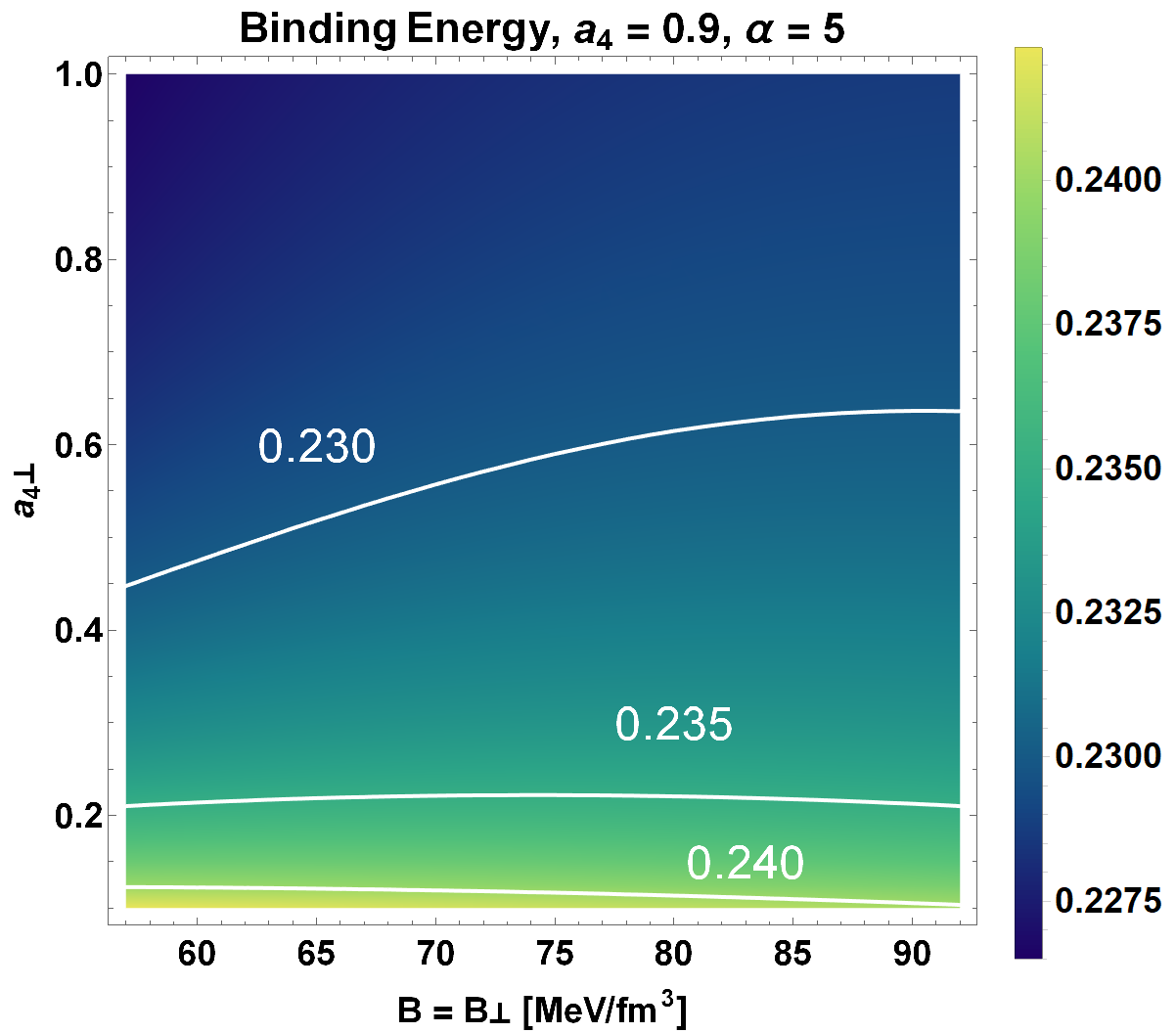}
    \caption{Binding energy profile for the full range of values of $B_{\perp}$ and $ a_4^{\perp}$. Each numerical values are given in Table \ref{table1}.}
    \label{binding}
\end{figure}

Finally, let us comment on the gravitational binding energy, $E_g$, shown in the four panels of Fig.~\ref{binding}. We have computed $E_g/M$ for $a_4 = 0.1, 0.5, 0.7, 0.9$, respectively, setting $\alpha=5~km^2$ for the entire allowed range for the bag constants $ B=B_{\perp}$, from $57~MeV/fm^3$ up to $92~MeV/fm^3$. Our numerical results demonstrate that the binding energy of anisotropic SQ stars takes its largest possible value at the bottom of the fourth panel, when the difference between $a_4$ and $a_4^{\perp}$ takes its maximal value.


\begin{table}[ht]
    \centering
    \begin{tabular}{|c|c|c|c|c|c|c|c|}
    \hline\hline
    $B = B_{\perp}$ & $a_4$ &  $a_4^{\perp}$  & $M_{\text{max}}$ &  $R$ & $\epsilon_{\text{c}}$ & $2MG/Rc^{2}$ & $Z_{\rm surf}$ \\
    (MeV/fm$^3$) &        &              & $(M_{\odot})$ & (km) & (MeV/fm$^3$) &  &  \\
    \hline
        57 & 0.2 & 0.07 & 2.17 & 10.46 & 1.58 $\times 10^{3}$ & 0.611 & 0.603 \\
           &     & 0.1 & 2.11 & 10.35 & 1.58 $\times 10^{3}$ & 0.601 & 0.583 \\
           &     & 0.3 & 1.99 & 10.15 & 1.52 $\times 10^{3}$ & 0.577 & 0.538 \\
           &     & 0.5 & 1.95 & 10.08 & 1.52 $\times 10^{3}$ & 0.571 & 0.526 \\
           &     & 0.7 & 1.93 & 10.04 & 1.52 $\times 10^{3}$ & 0.567 & 0.520 \\
           &     & 0.9 & 1.92 & 10.01 & 1.52 $\times 10^{3}$ & 0.565 & 0.516 \\
    \hline
        57 & 0.7 & 0.07 & 2.37 & 11.10 & 1.48 $\times 10^{3}$ & 0.631 & 0.645 \\
           &     & 0.1 & 2.31 & 10.99 & 1.48 $\times 10^{3}$ & 0.621 & 0.624 \\
           &     & 0.3 & 2.18 & 10.79 & 1.41 $\times 10^{3}$ & 0.596 & 0.573 \\
           &     & 0.5 & 2.14 & 10.76 & 1.35 $\times 10^{3}$ & 0.586 & 0.554 \\
           &     & 0.7 & 2.12 & 10.72 & 1.35 $\times 10^{3}$ & 0.583 & 0.548 \\
           &     & 0.9 & 2.11 & 10.69 & 1.35 $\times 10^{3}$ & 0.580 & 0.544 \\
    \hline
        70 & 0.5 & 0.07 & 2.13 & 9.94 & 1.84 $\times 10^{3}$ & 0.632 & 0.649 \\
           &     & 0.1 & 2.08 & 9.89 & 1.78 $\times 10^{3}$ & 0.621 & 0.624 \\
           &     & 0.3 & 1.98 & 9.68 & 1.78 $\times 10^{3}$ & 0.602 & 0.585 \\
           &     & 0.5 & 1.95 & 9.62 & 1.78 $\times 10^{3}$ & 0.596 & 0.573 \\
           &     & 0.7 & 1.93 & 9.59 & 1.78 $\times 10^{3}$ & 0.593 & 0.567 \\
           &     & 0.9 & 1.92 & 9.57 & 1.78 $\times 10^{3}$ & 0.591 & 0.563 \\
    \hline
        92 & 0.9 & 0.07 & 1.95 & 8.89 & 2.34 $\times 10^{3}$ & 0.647 & 0.683 \\
           &     & 0.1 & 1.91 & 8.82 & 2.34 $\times 10^{3}$ & 0.639 & 0.665 \\
           &     & 0.3 & 1.83 & 8.67 & 2.34 $\times 10^{3}$ & 0.623 & 0.629 \\
           &     & 0.5 & 1.81 & 8.63 & 2.34 $\times 10^{3}$ & 0.618 & 0.618 \\
           &     & 0.7 & 1.80 & 8.60 & 2.34 $\times 10^{3}$ & 0.615 & 0.612 \\
           &     & 0.9 & 1.79 & 8.59 & 2.34 $\times 10^{3}$ & 0.614 & 0.609 \\
    \hline
    \hline
    \end{tabular}
    \caption{Numerical outcomes for maximum gravitational mass and its corresponding radius for $B = B_{\perp}$  and $ a_4 \neq a_4^{\perp}$. All values are recorded for $\alpha = 5~km^2$.}
    \label{table1}
\end{table}


 Before we summarize our work, a final comment is in order here.
In a previous work \cite{Tangphati:2021tcy} some of us studied 
anisotropic quark stars within Einstein-Gauss-Bonnet is five dimensions. In both works the stars are non-rotating and electrically neutral, and also the equation-of-state for quark matter is the same. The field equations are formally the same in both works, and so is the total number of free parameters (e.g. the Gauss-Bonnet coupling and the parameters that enter into the EoS). There are, however, a few differences between the present work and the previous work \cite{Tangphati:2021tcy}, and those are the following: First of all, in four dimensions the Newton's constant, $G$, and the Planck mass, $m_{pl}$, are known, whereas in higher dimensions $G_D$ and $M_D$ are essentially unknown. In addition to that, although the field equations for the metric tensor are formally the same, in the structure equations the numerical pre-factors and the powers of $r$ change as one moves from four-dimensional to higher dimensional space-times. This results in slightly different quantitative numerical results, although qualitatively speaking the figures preserve the main features, and exhibit a similar behaviour in both cases, namely in $D=4$ and $D=5$.

\section{Summary and Discussion }\label{sec7}

In this work we have analyzed non-rotating quark stars assuming an interacting EoS for strange matter phase and allowing anisotropy in the pressure. The EoS (\ref{Prad1}) strictly depends on the bag constant, $B$, and the interaction parameter $a_4$, while it is reduced to the simplest MIT bag model when the quark mass $m_s$ is set to zero. In particular, we have considered the mass-radius relations for QSs in the framework of recently proposed $4D$ Einstein-Gauss-Bonnet (EGB) gravity. The generalized TOV equations are solved numerically to obtain the mass-radius $(M-R)$ relation as well as some other physical properties of the star. We found that the $(M-R)$ profile strictly depends on the three parameters i.e., two parameters coming from the EoS under consideration, the third one being
the GB coupling  constant $\alpha$, respectively. Studying the impact of the coupling constant $\alpha$, we see that the maximal mass increases by increasing this parameter. This trend is similar to the case of $5D$ EGB gravity, see Ref \cite{Tangphati:2021tcy}. But,  the maximum mass in our case can exceed $2 M_{\odot}$ limit at sufficiently small positive values of $\alpha$, where we see a clear deviation from $5D$ EGB gravity \cite{Tangphati:2021tcy}.

\smallskip

Next, we examined the effects of pressure anisotropy regime is attainable for two sets of solution (i) $ a_4 = a_4^{\perp}$ when $B \neq B_{\perp}$ and  (ii) $ a_4 \neq a_4^{\perp}$ when $B = B_{\perp}$, respectively. Then we varied the tangential components, $a_4^{\perp}$ and $ B_{\perp}$, while the radial components, $a_4$ and $B$, are being fixed. Table \ref{table1} summarizes the main aspects of the models used in this work, and Fig. \ref{Total_MR_fix_B} shows some associated $M-R$ relations for QSs. In the left panel of Fig. \ref{Total_MR_fix_B}, we see the possibility for the existence of stars with a mass $M > 2 M_{\odot}$ for $ a_4 = 0.7, ~  a_4^{\perp} = 0.2$ when $\alpha \geq 0$, with $\alpha = 0$ corresponding to the GR solution. For the lowest value of bag parameter, $B = 57~\,{\rm MeV}/{\rm fm}^{3}$, we find the maximum masses of compact stars. Therefore, the compact massive QS can be explained in this specific version of $4D$ EGB gravity.

\smallskip

We have also shown that the parameter dependence diagrams related to $M-\epsilon_c$, compactness and binding energy, respectively. For a sequence of stars modeled by the same EoS, we studied the stability
criterion by the following inequality $\frac{\partial M(\epsilon_c)}{\partial \epsilon_c}$ $\lessgtr0$.  
As we can see, from the results in Fig. \ref{Total_ME_fix_B} that $a_4 = a_4^{\perp}$ gives more stable solution than $B = B_{\perp}$. Though it is a necessary condition for recognizing stable configurations. We have also studied the effects of the parameters on the compactness and binding energy in Figs. \ref{Total_Com_fix_B} and \ref{binding}, respectively.  For a fixed values of bag constant, we found that compactness and binding energy are decreasing monotonically when increases the value of $a_4^{\perp}$ from  $0.07$ to $0.9$, see Table \ref{table1}. Our findings indicate that less interacting quarks have larger binding energy. In addition to this, the obtained solution is less compact and less stable compared to $5D$ EGB theory.

\smallskip

Therefore, we may conclude that the model proposed and analysed in detail is both viable and realistic, since it passes a series of theoretical and observational constraints, while at the same time it is able to describe the basic properties of quark matter associated to compact objects.


\section*{Acknowlegements}

We wish to thank the anonymous reviewer for comments and suggestions. The author T.~T would like to thank the financial support from the Science Achievement Scholarship of Thailand (SAST). A. Pradhan thanks to IUCCA, Pune, India for providing facilities under associateship programmes. The author G.~P. thanks the Fun\-da\c c\~ao para a Ci\^encia e Tecnologia (FCT), Portugal, for the financial support to the Center for Astrophysics and Gravitation-CENTRA, Instituto Superior T\'ecnico, Universidade de Lisboa, through the Project No.~UIDB/00099/2020 and No.~PTDC/FIS-AST/28920/2017. 


\end{document}